\documentclass[seceq]{ptptex}
\title{%        %You can use \\ for explicit line-break
Deterministic Chaos in Quantum Field Theory%
}

\author{%       %Use \scshape  for the family name
Viatcheslav \textsc{Kuvshinov}$^1$ and Andrei \textsc{Kuzmin}$^2$%
}

\inst{%         %Affiliation, neglected when [addenda] or [errata]
$1$-Institute of Physics, 220072 Belarus, Minsk, Scorina,68.
                    Tel. 375-172-84-16-28, fax: 375-172-84-08-79.
                    E-mail: kuvshino@dragon.bas-net.by

$2$-Institute of Physics, 220072 Belarus, Minsk, Scorina,68.
                    Tel. 375-172-84-16-28.
                    E-mail: avkuzmin@dragon.bas-net.by
                    }

\abst{%         %this abstract is neglected when [addenda] or [errata]
We discuss the necessity and demonstrate the validity of introduction the notion of
deterministic chaos in quantum field theory. Brief review of the existing approaches
to this problem is given. We compare proposed chaos criterion for quantum fields with
existing ones. Its consequences in particle physics are also discussed. }

\begin{document}

\maketitle

\section{Introduction}\label{Introduction}

Chaos phenomenon attracts much attention in various fields of natural sciences, from
celestial mechanics~\cite{Sagdeev}( pp.349-359) and behavior of social systems
\cite{Shuster} to quantum mechanics \cite{Zaslavsky} and atomic physics \cite{ion}.
Though the understanding of the reasons and regimes of chaotic behavior in classical
(Hamiltonian) systems is already achieved in main~\cite{Lihtenberg}, problems
connected with properties, manifestations and even definition of deterministic chaos
in quantum world are still basically open~\cite{Bunakov,PLA}. Prevalence of chaotic
motions in the majority of natural phenomena explains the fundamental and applied
significance of its study. Particulary, chaotic behavior is intrinsical to classical
gauge fields in modern theories of particle interactions \cite{Savvidy,N,we2002} as
well it plays crucial role for quantum tunneling control, which is of practical
importance \cite{Tunneling}.

In this paper we discuss chaos phenomenon in the context of quantum field theory
(QFT). We premise the consideration of our main topic by some motivations.
Investigation of the classical gauge field dynamics from the viewpoint of chaos was
started in a few years after recognition of gauge theories based on $SU(3)$ and
$SU(2)\otimes U(1)$ groups as theories of strong and electroweak interactions
respectively. Nonabelian gauge fields form complicated nonlinear system with large
number of degrees of freedom, which behavior demonstrates a lot of regimes when
parameters vary. Besides the investigation of classical gauge fields to be interesting
from the viewpoint of nonlinear dynamics, it is also important from the viewpoint of
particle physics (particulary QCD), where large progress is achieved in understanding
of perturbative phenomena, but comparatively little is known about the
non-perturbative evolution of quarks and gluons. Namely, the confinement phenomenon
responsible for absence of free colored objects is waiting for its explanation
(however a lot of results is obtained in this direction, for review see
Ref.~\cite{Nonpert}). The problem is that there are no exact methods for description
of non-perturbative evolution of quantum fields, when coupling between them is not too
small (there is no small parameter) and perturbative expansions can not be applied.
The first step for understanding of quantum dynamics in this coupling regime is the
investigation of the classical behavior of fields in this region. However, even
analysis of classical dynamics appears to be a hard problem, because of infinite
number of degrees of freedom of field system. Its reduction to finite one is needed in
order existing methods to be applied. The easiest way is to consider model field
system with small number of freedoms. It is achieved by imposing "by hands" some
additional restrictions on the behavior of fields. For example it can be the condition
of spatial homogeneity of gauge fields, which was used by many authors~\footnote{In
the certain Lorenz frame there exist exact spatially homogeneous solutions of gauge
field equations of motion. They were primarily obtained in the Ref.~\cite{SHS}.}, for
instance \cite{Savvidy,we2002,weJNMP} (see references therein). On the examples of
spatially homogeneous model field systems it was demonstrated that the behavior of
nonabelian classical gauge fields is inherently chaotic \cite{Savvidy}. Existing
numerical methods let to consider more realistic model of continuous in space-time
gauge field system possessing thousands of degrees of freedom, namely, gauge fields on
a lattice. Maximal Lyapunov exponent \cite{MaxL} and total Lyapunov spectrum
\cite{TotalL} were numerically determined. Lattice simulations straightly demonstrated
that the dynamics of gauge fields on a lattice is chaotic. Thus one can expect that
real classical gauge field system demonstrates chaotic behavior, however some
differences between gauge field system on a lattice and in continuous space-time have
been noted \cite{N2}.

Chaotic behavior of classical gauge fields does not directly indicate how should chaos
phenomenon become apparent in quantum case and what is deterministic chaos in quantum
field theory. However there are evidences of chaotic behavior in modern quantum field
theories such as QCD, namely signs of chaos in branching processes \cite{Branch}. This
approach do not refer to the chaotic behavior of classical gauge fields and thus
represents alternative way to the notion of chaos in quantum field theory. The
question about its relation with chaotic dynamics of classical gauge fields is open.

The aim of this work is to present our view on the main problems lying on the path
leading to the notion of chaos in quantum field theory and to present our approach to
this problem. We consider proposed by us chaos criterion for quantum fields \cite{PLA}
in the context of existing results obtained in this direction. Also we discuss
possible manifestations of deterministic chaos in quantum field theory. Particulary in
connection with confinement phenomenon.

The paper is organized as follows. In the Sec.\ref{2} we review existing quantum chaos
criteria from the viewpoint of their applicability to quantum field theory. Also we
discuss our chaos criterion for quantum fields. Its correspondence with the notion of
chaos for classical fields is qualitatively demonstrated. In the Sec.\ref{3} we
provide further justification of proposed chaos criterion. Its correspondence with
existing quantum chaos criteria and some consequences are discussed also. Conclusion
can be found in the Sec.\ref{4}.

\section{Chaos criterion for quantum fields}\label{2}

Necessity of probabilistic (or statistical) description of physical systems was
primarily realized when behavior of the systems with extremely large number of degrees
of freedom was studied (statistical mechanics). Mainly in the framework of classical
mechanics of Hamiltonian systems it was understood that chaotic behavior of mechanical
systems is determined by its local instability rather then a large number of degrees
of freedom \cite{Sagdeev}. Chaotic behavior is prevalent in dynamical regimes of
classical mechanical systems. Regular motion is rather exceptional \cite{Zaslavsky}.
It was noticed in the Ref.~\cite{Bunakov} that since classical mechanics is a limiting
case of more fundamental quantum mechanics then basic reasons for classical chaotic
motion one has to search in quantum world. Exponential divergence of neighbor
trajectories (local instability), mixing and other attributes of classical chaos have
to be considered as consequences of quantum picture. At the moment the opposite
approach dominates. One considers systems with classically regular, mixed or pure
chaotic motion and compares properties of their quantum analogues in order to find
some differences and special features which are refereed as "quantum chaos". Large
progress is achieved in this direction. Particulary, universal classes of spectral
fluctuations for quantum systems with chaotic behavior in the classical limit have
been obtained~\cite{Robnic}. Any other approach to the problem of quantum chaos has to
be in agreement with these results.

Spectral properties of classically regular or chaotic quantum systems were explained
by Gutzwiller's periodic orbit theory~\cite{Gut}. The trace formulae~\cite{Gut}
relates quantum mechanical Green function with the spectral density of states. It
provides the bridge between the formulation of quantum chaos in spectral terms and its
path integral formulation, which is more convenient for the extrapolation to quantum
fields.

For the role of "fundamental" chaos criterion valid for quantum system the value of
the system's symmetry violation was proposed \cite{Bunakov}. The parameter
quantitatively characterizing symmetry violation for Hamiltonian systems was
introduced and correspondence with classical chaos criteria was checked on the
examples of Hennon-Heiles system and diamagnetic Kepler problem
\cite{Bunakov,Bunakov2}. The language of symmetry is the universal language of quantum
(not only) physics. However, realization of symmetric approach, proposed in
\cite{Bunakov,Bunakov2} and well justified in quantum mechanics and nuclear physics,
from our point of view, can not be extrapolated in straightforward way for quantum
fields. The reason is the absence of evident relativistic covariance and emphasized
role of the energy. But the principle of using the symmetry of the system as the
measure of its regularity proposed in the Ref.~\cite{Bunakov} is fundamental and,
undoubtedly, it will be claimed in some form in quantum field theory.

The notion of chaos in quantum field theory can also be introduced in terms of
eigenvalue spectrum of the lattice Dirac operator \cite{D1}. Particulary, it was
demonstrated that the nearest-neighbor spacing distribution for the eigenvalue
spectrum of the staggered Dirac matrix in quenched QCD on a lattice agrees with the
Wigner surmise of random matrix theory \cite{D2}.

Quantum chaos criterion formulated in classical terms was proposed in the
Ref.\cite{Eff}, where quantum system was defined as chaotic if its renormalized action
provides {\it classically} chaotic dynamics (when minimal action principle is
applied). The advantages of this approach are the obvious correspondence with
classical chaos criteria (if Plank constant equals zero then renormalized action turns
into classical action) and relativistic invariance when this criterion is extrapolated
to quantum field theory.

We proposed to give axiomatic formulation of relativistic and gauge~\footnote{The
reason is that all modern theories of particle interactions are gauge theories.}
invariant chaos criterion in quantum field terms only~\cite{PLA}. No doubt that some
classical motivations which brought us to the formulation of the chaos criterion had
to be used. Further justification of the proposed criterion and the check of its
correspondence with the notion of classical chaos are to be done.

Now we give some qualitative arguments which bring us to formulation of chaos
criterion in quantum field theory \cite{PLA}. From statistical mechanics and ergodic
theory it is known that chaos in classical systems is a consequence of the property of
mixing \cite{Lihtenberg}. Mixing means rapid (exponential) decrease of correlation
function with time \cite{Sagdeev}. In other words, if correlation function
exponentially decreases then the corresponding motion is chaotic, if it oscillates or
is constant then the motion is regular \cite{Shuster}. We expand criterion of this
type for quantum field systems.  All stated bellow remains valid for quantum
mechanics, since mathematical description via path integrals is the same.

For quantum field systems the analogue of classical correlation function is the
two-point connected Green function
\begin{equation}\label{green}
G_{ik}(x,y)= -\frac{\delta^{2}W[\vec{J}]}{\delta J_{i}(x)
 \delta J_{k}(y)}\left.\right|_{\vec{J}=0}.
\end{equation}
Here $W[\vec{J}]$ is the generating functional of connected Green functions, $\vec{J}$
are the sources of the fields, $x$, $y$ are 4-vectors of space-time coordinates.
\par
Thus we formulate chaos criterion for quantum field theory in the following form
\cite{PLA}:
\par
a) If two-point Green function (\ref{green}) exponentially (or faster) goes to zero
when the distance between its arguments goes to infinity then system is chaotic.
\par
b) If it oscillates and/or slower then exponentially goes to zero in this limit then
we have regular behavior of quantum system.

Obviously that proposed chaos criterion is relativistic and gauge invariant and
formulated in terms of quantum field theory only. Also it has direct physical sense,
namely, if the propagator decays exponentially or faster, then this case corresponds
the chaotic behavior of quantum field system. In opposite case the dynamics is
regular. Particulary it is seen that dynamics of free quantum fields is always regular
as it should be~\footnote{Decreasing of the free field propagator is slower then
exponential.}.

Our aim is to demonstrate the correspondence between proposed definition of chaos for
quantum fields and chaotic behavior of classical fields in the semiclassical limit
$\hbar \rightarrow 0$. Providing it we justify our quantum chaos criterion initially
postulated.

Primarily, we check the agreement between classical criterion of local instability and
formulated quantum chaos criterion in framework of quantum mechanics. After this we
sketch the prove for QFT. Particulary, we qualitatively justify the correspondence
between the quantum chaos criterion and the notion of chaos for classical fields.

For non-relativistic quantum mechanics the problem is still too complicated to be
solved analytically in the closed form. Recent results concerning the calculation of
the propagator and Green functions in this case see in Ref.\cite{QMGreen}. We consider
the classical systems with Hamiltonian having the form:
\begin{equation}\label{Hamiltonian}
  H=\frac{1}{2}\vec{p}^{2}+V(\vec{q})\quad,\quad
\vec{p}=(p_{1,\ldots,}p_{N})\;;\;\vec{q}=(q_{1,\ldots ,}q_{N}), \quad N>1
\end{equation}
Here $N$ is the (arbitrary) finite number of degrees of freedom.

We consider the case of constant eigenvalues of the classical stability matrix (matrix
form of the Jacobi-Hill operator, controlling the linear stability around the
classical orbit). The case of non-constant eigenvalues will be qualitatively
considered bellow. Stability matrix for the Hamiltonian (\ref{Hamiltonian}) is given
by the following expression \cite{PLA}:
\begin{equation}
G\equiv\left(
\begin{array}
[c]{cc}%
0 & I\\
-\Sigma & 0
\end{array}
\right); \quad \Sigma\equiv\left(  \frac{\partial^{2}V}{\partial q_{i}\partial
q_{j}}\left| _{\vec{q}_{0}}\right.  \right).
\end{equation}
Here $I$ is the $N \times N$ identity matrix. Eigenvalues of the stability matrix $G$
are functions of configuration space point $\vec{q}_0$. They determine evolution of
small deviations between two neighbor trajectories in the phase space. Namely solution
of linearized Hamilton equations valid in the vicinity of considered configuration
space point has the form:
\begin{equation}\label{solution}
\left(
\begin{array}
[c]{l}%
\delta\vec{q}(t)\\
\delta\vec{p}(t)
\end{array}
\right)  =\sum_{i=1}^{2N}C_{i}\exp\left\{  \lambda_{i}t\right\} \left(
\begin{array}
[c]{l}%
\delta\vec{q}(0)\\
\delta\vec{p}(0)
\end{array}
\right).
\end{equation}
Here $\lambda_{i}=\lambda_{i}(\vec{q}_{0})$ are eigenvalues of the stability matrix
$G$. And $\{C_{i}\}$ is a full set of projectors. From (\ref{solution}) it is seen:
\par
a) If there is $i$ such as $\operatorname{Re}\lambda _{i}>0$ then the distance between
neighboring trajectories grows exponentially with time and motion is locally unstable.
According Liouville's theorem stretching of phase space flow in one direction
($\operatorname{Re} \lambda_{i} >0$) is accompanied by its compression in other
direction (directions) in order to keep phase space volume constant. That means the
existence of $\operatorname{Re} \lambda_{j} < 0$. Thus for local instability of motion
we can demand existence of $\operatorname{Re}\lambda _{k} \neq 0$.
\par
b) If for any $i=\overline{1,2N}\quad$ $\operatorname{Re}\lambda_{i}=0$ \ then there
is no local instability and the motion is regular.

In the case of {\it constant} increments of local instability $\{\lambda_i\}$
two-point connected Green function (\ref{green}) in the semiclassical limit can be
represented in the form \cite{PLA}:
\begin{equation}\label{MainGreen2}
 G_{i}(t_{1},t_{2})=\frac{i}{2}\operatorname{Re}\left( \frac{e^{-\lambda_{i}(t_{1} -
 t_{2})}}{\lambda_{i}} \right), \quad t_{1}>t_{2}.
\end{equation}
From the expression (\ref{MainGreen2}) it is seen:
\par
a) If classical motion is locally unstable (chaotic) then according criterion stated
above there is real eigenvalue $\lambda_{i}$. Therefore Green function
(\ref{MainGreen2}) exponentially goes to zero for some $i$ when
$(t_{1}-t_{2})\rightarrow +\infty$. Opposite is also true. If Green function
(\ref{MainGreen2}) exponentially goes to zero under the condition
$(t_{1}-t_{2})\rightarrow +\infty$ for some $i$, then there exists real eigenvalue of
the stability matrix and thus classical motion is locally unstable.
\par
b) If all eigenvalues of the stability matrix $G$ are pure imaginary, that corresponds
classically stable motion, then in the limit $(t_{1}-t_{2})\rightarrow +\infty$ Green
function (\ref{MainGreen2}) oscillates as a sine. Opposite is also true. If for any
$i$ Green functions oscillate in the limit $(t_{1}-t_{2})\rightarrow +\infty$ then
$\{\lambda_{i}\}$ are pure imaginary for any $i$ and classical motion is stable and
regular.
\par
Thus we have demonstrated for any finite number of degrees of freedom that proposed
quantum chaos criterion coincides with classical criterion of local instability in the
semiclassical limit of quantum mechanics in the case when increments of local
instability are constant (corresponding principle).

In the case of non-constant $\lambda$s the calculation of two-point connected Green
function in the whole range of variation of its arguments is several orders more
complicated problem then in the case of constant ones. The condition for Green
function to be finite in the limit of infinite distance between its arguments forced
us to eliminate exponentially growing item from the expression (\ref{MainGreen2}).
However, it is not so in general case, when we can consider increments of instability
as constants just in small region around the considered point of configuration space.
Therefore we can not demand the elimination of exponentially growing item and the
expression for two-point connected Green function valid in sufficiently small region
of configuration phase space is:
\begin{equation}\label{TrueGreen}
  G_i(t_1,t_2) = D_1^{(i)} e^{\lambda_i (t_1-t_2)} + D_2^{(i)} e^{-\lambda_i
  (t_1-t_2)},
\end{equation}
where $D_1^{(i)}$, $D_2^{(i)}$ are arbitrary constants and $t_1-t_2$ is assumed to be
sufficiently small. Thus for non-constant $\lambda$s we can describe {\it local}
behavior of Green function, but we are not able to predict its {\it global} behavior
that is needed for proposed chaos criterion to be applied.

Above we provided justification of proposed chaos criterion for non-relativistic
quantum systems with any finite number of degrees of freedom which can be considered
as QFT in $0+1$ dimensions. However quantum mechanics is just formally similar to
quantum field theory. Physically they are different systems. Namely, quantum
mechanical systems considered above have any but {\it finite} number of degrees of
freedom whereas quantum fields possess {\it infinite} number of freedoms and have to
be considered in the space-time which dimension is larger then one. Rapid decreasing
of the Green function (\ref{green}) for quantum mechanical system means localization
in configuration space~\footnote{The notion of configuration space can be introduced
since we consider semiclassical limit.} of the system while for quantum fields
localization in space-time is required. Why did we consider quantum mechanics in this
case? The answer is that for field system localization in its configuration space is
the necessary condition for the space-time localization. This provides us the way for
further justification of the proposed chaos criterion.

Now we qualitatively justify the correspondence between proposed quantum chaos
criterion in semiclassical limit and chaotic behavior of classical fields in the
general case of arbitrary field system. Some specification are possible in concrete
cases. But we believe that the sketch of prove remains the same.

We state that chaotic behavior of classical fields leads to exponential or faster
decreasing of the Green function (\ref{green}) in the semiclassical limit. Without
loss of generality we can consider the system consisting from the fields and/or field
components $\varphi_a (x)$. Here $x$ denotes the space-time 4-vector, $a$ enumerate
fields and field components. Let us consider the field system on a spatial lattice and
leave time continuous. We break relativistic invariance in the manner widely used in
the lattice QFT \cite{kogut}. Field functions $\varphi_a (\vec{x}_j , t)$ in discrete
space are generalized coordinates of the field system in the classical case and play
the role of Heisenberg generalized coordinate operators after the secondary
quantization. They obey usual quantum mechanical commutation relations:
\begin{equation}\label{CommRel}
 \left[\varphi_a (\vec{x}_i , t), \varphi_b (\vec{x}^{\prime}_j , t)\right] =
 i\hbar_{eff} \delta_{ab} \delta_{ij}.
\end{equation}
Here $\hbar_{eff}$ is the effective Plank constant. Hamiltonian of the field system on
a spatial lattice has the form (\ref{Hamiltonian}). Chaotic behavior of classical
fields means uncorrelated behavior in different space points and exponential or even
faster decay of the classical correlation function with time. That implies locally
unstable motion in the phase space of the field system. According the results of
Ref.\cite{PLA} it leads to exponential decreasing of the Green function (\ref{green})
for constant eigenvalues of the stability matrix (increments of local instability).
The general case of non-constant increments of local instability can be considered
only qualitatively. There are indications originated from quantum mechanics that
localization in the phase space appears in the presence of classical chaos, however
the question about the relation between classical chaos and quantum dynamical
localization is still open \cite{Robnic}.

Qualitative justification of proposed chaos criterion in the general case looks as
follows. Field system on a lattice built above is a Hamiltonian system with large
number of degrees of freedom and bounded motion in its phase space (corresponding
boundary conditions are assumed). In semiclassical regime which is considered there
exists break (or Heisenberg) time $\tau_H$. It determines the time scale up to which
quantum dynamics follows classical one. Break time decreases when effective Plank
constant $\hbar_{eff}$ determined by (\ref{CommRel}) increases. If Heisenberg time
becomes less then the classical diffusion time $\tau_D$ needed the system to cover all
available phase space ($\tau_H < \tau_D$) then the field system is dynamically
localized in the phase (therefore configuration) space (see the Ref.~\cite{Robnic}).
That is the necessary condition for the space-time localization required by the
proposed chaos criterion.

To demonstrate it consider localized in space field configuration. In the simplest
case localization means the field functions to be zero outside some finite hypersphere
in configuration space of the system. As well fields are assumed to take finite values
because of physical reasons (no singularities). Therefore, under the conditions stated
the space-time localization leads to the localization in the configuration space. That
was needed to prove. The question about the sufficient conditions is still open.

\section{Consequences of proposed chaos criterion}\label{3}

In this section we discuss consequences of deterministic chaos in QFT following from
the proposed chaos criterion (see the Sec.\ref{2}). Particulary, rapid (exponential or
faster) decrease of the propagator (\ref{green}) implies the system to be confined in
some region of the space-time. The same behavior of the propagator is required in
order to provide the regime of superlocalization~\cite{Simonov1} needed for
explanation of the confinement of colored objects (quarks and gluons) in
QCD~\cite{Simonov2}. Therefore deterministic chaos in QFT regarded in the framework of
our approach can be considered as the sufficient condition for the confinement
phenomenon to occur. Moreover the direct connection between the confinement of
particles and stochastic behavior of the background classical fields was
obtained~\cite{Simonov2}. This fact give additional argument~\footnote{It does not
prove the correspondence between proposed quantum chaos criterion and classical ones,
because the deterministic chaos is the sufficient condition for stochastic behavior
but it is not the necessary one.} in favour of the chaos criterion proposed in the
Ref.~\cite{PLA}. The condition of stochasticity of classical background non-abelian
gauge fields in the simplest case means fast enough decrease of the bilocal correlator
in the cluster expansion~\cite{Kampen} of the path ordered exponential in the
fundamental representation~\cite{Simonov3} (conditions needed for applicability of
non-abelian Stocks theorem~\cite{Simonov3} are assumed to be justified). Roughly, for
confinement of quarks and gluons to occur (with linear confining potential in
non-relativistic limit) existence of the finite correlation length for the classical
background non-abelian gauge fields is needed~\cite{Simonov2}. In the
Ref.~\cite{Simonov2} the stochastic assembly of background fields imposed "by hands"
was considered. We note that it can be realized by the classical chaotic solutions of
the field equations of motion. These configurations are found both as exact solutions
of gauge field equations~\cite{Savvidy,we2002} (see references wherein) and in lattice
simulations~\cite{N,TotalL}. They are not vacuum configurations and the condition of
(anti-) self-duality assumed in Refs.\cite{Simonov2,Simonov3} is broken. However they
realize the local minimum of the field action and therefore provide non-zero amplitude
for realization of the confinement mechanism. Moreover due to a large number of such
solutions their contribution can be essential and even exceed the contribution of
vacuum configurations, this is the question for further investigations.

One of possible applications of proposed chaos criterion in field theory is an
investigation of the stability of classical solutions with respect to small
perturbations of initial conditions. Of course, this does not directly imply chaos,
but advances us to it. To study the stability of certain classical solution of field
equations one has to calculate (for instance, using one loop approximation) two-point
Green function in the vicinity of considered classical solution.
\par
To demonstrate this, consider real scalar $\varphi^{4}$-field:
\begin{equation}\label{L}
L= \frac{1}{2} \left( \partial_{\mu} \varphi \right)^{2} - \frac{1}{2}m^{2}
\varphi^{2} - \frac{\lambda}{4!} \varphi^{4}.
\end{equation}
Here $\lambda>0$ is a coupling constant, $m^{2}$ is some parameter which can be larger
or less then zero. In both cases $\varphi=0$ is a solution of field equations.
Asymptotic of two-point Green function calculated in the vicinity of the classical
solution $\varphi=0$ in the zero order of perturbation theory is:
\begin{equation}\label{Lastf}
G(x,y)_{\widetilde{\rho \rightarrow \infty}} \rho^{-\frac{1}{2}}e^{im\sqrt{\rho}}.
\end{equation}
Here $\rho=(x-y)^{2}$ and we accept that 4-vector $x-y$ is inside the light cone
$(x^{0}-y^{0})>0$, in other words $\rho>0$. We can study the stability of considered
solution with respect to small perturbations. Expression (\ref{Lastf}) shows that we
have two different cases:
\par
a) Green function oscillates and slowly (non-exponentially) goes to zero when $\rho
\rightarrow \infty$. According proposed chaos criterion considered solution is stable.
Indeed, from (\ref{Lastf}) it follows that parameter $m$ is real in this case.
Therefore $\varphi=0$ is a stable vacuum state.
\par
b) Green function exponentially goes to zero in the limit $\rho \rightarrow \infty$.
From proposed chaos criterion it follows that $\varphi=0$ is an unstable solution.
That is true since from (\ref{Lastf}) one can see that parameter $m$ has to be pure
imaginary. It is known that in this case state $\varphi=0$ becomes unstable, two new
stable vacuums are appeared and we obtain spontaneous symmetry breakdown \cite{Higgs}.
\par
Thus for real scalar $\varphi^{4}$-field spontaneous symmetry breakdown and
degeneration of vacuum state can be regarded as signatures of quantum chaos. This
relates our approach with the symmetry approach of Bunakov~\cite{Bunakov,Bunakov2}
(see discussion in the Sec.~\ref{Introduction}). Namely, on the particular example we
demonstrated proposed chaos criterion to lead to the ground state symmetry violation.

\section{Conclusion}\label{4}

In this work we have demonstrated the necessity and substantiated the validity of
introduction the notion of deterministic chaos in QFT. We briefly reviewed existing
approaches to this problem. We continued the justification of the chaos criterion for
quantum fields proposed by us in our earlier papers~\cite{PLA,we2002}. Particulary, we
demonstrated semi-qualitatively that exponential (or faster) decreasing of the
two-point connected Green function (\ref{green}) is the sufficient condition for
chaotic behavior of the fields in the classical limit. Our qualitative arguments are
supported by the results obtained in Refs.~\cite{Simonov1,Simonov2,Simonov3} in
connection with confinement problem in QCD. We also conjectured that confinement of
quarks and gluons can be provided by classical chaotic solutions of Yang-Mills
equations. Relation between chaos in QFT systems and their symmetry violation has been
discussed on a particular example of the $\lambda\varphi^4$ field system.

We gave a brief review of the problems standing on the way to the understanding the
nature and consequences of deterministic chaos in QFT. Particulary, the further
justification of the proposed chaos criterion is needed, as well its consequences has
to be clarified. Connection with the longstanding problems of particle physics, such
as confinement problem in QCD and other non-perturbative phenomena, has to urge
forward the investigation of deterministic chaos in QFT.

%\section*{Acknowledgements}
%We would like to thank ...........

%\appendix
%\section{First Appendix} %Empty argument \section{} yields `Appendix'.
%
%\section{Second Appendix}

\end{document}